\documentclass[%
 reprint,
showpacs,preprintnumbers,
 amsmath,amssymb,
 aps,
prl
]{revtex4-1}
\usepackage{ulem}
\usepackage{amssymb}
\usepackage{amsmath}
\usepackage{lipsum}
\usepackage{comment}
\usepackage{color}

\newcommand{\tcb}{\textcolor{black}}

\newcommand{\ti}{\tilde}

\usepackage{graphicx}
\usepackage{dcolumn}
\usepackage{bm}

\begin{document}

\preprint{APS/123-QED}

\title{Epigenetic Ratchet: Spontaneous Adaptation via Stochastic Gene Expression}

\author{Yusuke Himeoka}
\affiliation{Department of Basic Science, University of Tokyo, Komaba, Meguro-ku, Tokyo 153-8902, Japan\\
Center for Models of Life, Niels Bohr Institute, University of Copenhagen,
Copenhagen, Denmark}
\author{Kunihiko Kaneko}%
 \email{kaneko@complex.c.u-tokyo.ac.jp}
\affiliation{Research Center for Complex Systems Biology, Universal
Biology Institute, University of Tokyo, 3-8-1 Komaba, Tokyo 153-8902, Japan} 

\date{\today}

\begin{abstract}
Adaptation mechanism of cells on the basis of stochastic gene expression and epigenetic modification is proposed. From marginally stable states generated by epigenetic modification, a gene expression pattern that achieves greater cell growth is selected, as confirmed by simulations and analysis of several models. The mechanism does not require any design of gene regulation networks and is shown to be generic in a stochastic system with marginal stability. General relevance of the mechanism to cell biology is also discussed.

\end{abstract}

\maketitle

\begin{figure}[htbp]
\begin{center}
\includegraphics[width = 85 mm, angle = 0]{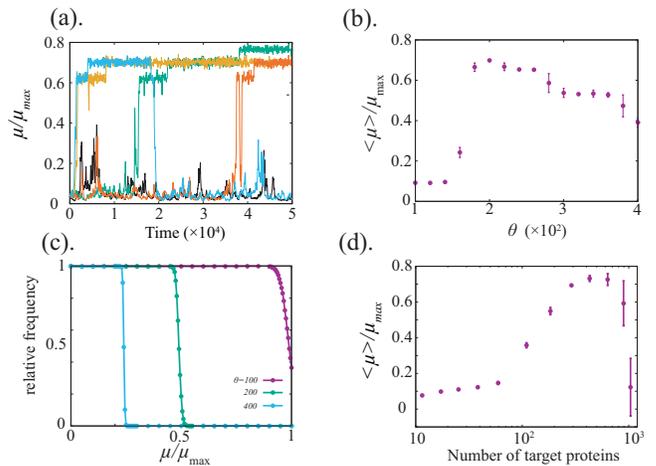}
     \caption{Simulation of our model via adaptation by stochastic reactions. (a) Time courses of the growth rate relative to its maximum value for $\theta=0$ (black curve) and $200$ (other curves). Although the time course with $\theta=0$ fails to increase the growth rate, other time courses with $\theta=200$ sequentially increase the growth rate. (b) The average growth rate as a function of $\theta$. (c) Relative frequency of a modification event plotted against the growth rate $\mu/\mu_{\rm max}$. The relative frequency is obtained by dividing by that at $\mu=0$. Data resulting from different values of $\theta$ are plotted with different colors. The frequency is obtained for a given growth rate at the moment. (d) Dependence of the average growth rate on the number of molecules in the reaction system. The number of reacting molecules is changed by changing the volume of the system $\Omega$. For (b) and (c), each point is computed by averaging $50$ samples of a time-averaged growth rate. Error bars indicate standard deviation. The samples were obtained by simulations with different seeds of random numbers, whereas $E_t$ and $P_t$ were set to the attractor value initially. The growth rate is averaged from $t=10^5$ to $t=10^6$. Parameters are set as $v_s=v_t=10^2,d_s=10^{-2},d_t=10,l_\pm=10^{-3},\mu_{\rm max}=1,P_t^*=15 $, and $\theta=200$ for (d). Different $\Omega$ values are employed to reduce the computation time, the respective values are (a) $20$, (b) $50$, and (c) $10$.}
    \label{fig:simple}
  \end{center}
\end{figure}

\begin{figure}[htbp]
\begin{center}
\includegraphics[width = 85 mm, angle = 0]{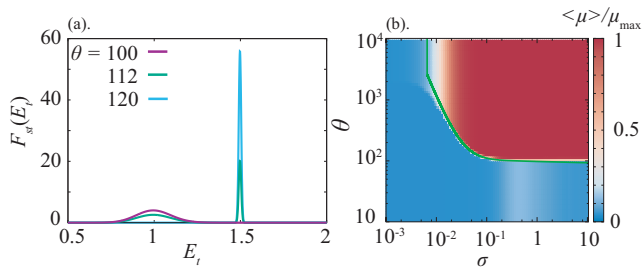}
     \caption{A distribution and the average growth rate obtained from the Fokker--Planck equation. (a) A stable distribution of $E_t$ plotted for $\theta=100,112$ and $120$. The peak of distribution moves from the steady-state value of Eq.(\ref{eq:reduce_det}) without a noise term to the neighbor of the target value $E_t^*$, as $\theta$ increases. (b) The average growth rate plotted as a function of the noise amplitude $\sigma$ and threshold $\theta$. \tcb{The green line is the boundary which separates the adaptive and non-adaptive region.}  Parameter values are $l_+=l_-=10^{-2},v=10,d_t=1,\ti{\beta}=5.0,\mu_{\rm max}=1,P_t^*=15$, and $\sigma=0.1$ for (a).}
    \label{fig:FP}
  \end{center}
\end{figure}

\begin{figure}[htbp]
\begin{center}
\includegraphics[width = 50 mm, angle = 0]{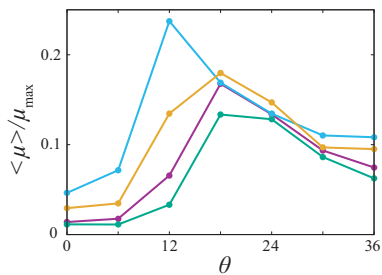}
     \caption{The averaged growth rates computed from three metabolic networks are overlaid. The growth rates are enhanced in the intermediate value $(10\sim20)$ of $\theta$. Each point is computed by averaging $5$ samples of a time-averaged growth rate. Details of the simulations, and parameter values are given in {\it Supplement}}
    \label{fig:network}
  \end{center}
\end{figure}

\tcb{Cells can adapt to a variety of environmental conditions to achieve a higher growth state. This adaptation is attained by changing the gene expression pattern, with which cells can grow upon different environmental conditions.} According to the seminal work by Jacob and Monod \cite{jacob1961genetic}, such a response of a gene expression pattern is implemented through a signal transduction network that translates environmental conditions to the actions on the promoter of the genes responsible for adaptation. In spite of the importance of such an adaptation mechanism, there remains a question to be addressed: In order to adapt to a variety of environmental conditions, cells have to prepare signal transduction networks corresponding to all of them, for which a huge variety of chemicals and genes would be needed, which may exceed the capacity of a cell. Besides, with such a mechanism, cells would not be able to adapt to the environmental changes that they have never experienced.\\
Indeed, recent experiments suggest that cells can adapt even to a novel, unforeseen environment by changing a gene expression pattern \cite{kotte2010bacterial,braun2015unforeseen,kashiwagi2006adaptive,stern2007genome,stolovicki2006synthetic}. \tcb{ In this case there is no room to evolve signal networks for such environment. In addition, some experiments have revealed that bacterial cells with an artificial gene network without corresponding signal transduction networks can show an adaptive response \cite{kashiwagi2006adaptive,tsuru2011adaptation,shimizu2011stochastic,tan2009emergent}.\\}
\tcb{Accordingly, a possible theoretical mechanism for the selection of the adaptive state has been proposed \cite{kashiwagi2006adaptive,furusawa2008generic,furusawa2013epigenetic}: When a cell has multiple attractors with different growth rates, the attractor with faster growth is less perturbed by stochasticity in gene expression dynamics, and cells tend to be kicked out from slow-growth attractors and attracted to those with faster growth. For this attractor selection mechanism by noise to work, however, gene regulation networks (GRNs) have to be prepared to have multiple attractors allowing for a higher growth rate under given environmental conditions. This situation will strongly limit the applicability of the mechanism. Can selection of adaptive states with higher growth rates be generally achieved without designing the GRN to have multistability? \\}
Indeed, epigenetic modifications based on several factors, including DNA or histone modification, and their interplay with higher-order chromatin structure are known to play key roles to modify the gene expression patterns of cells \cite{jaenisch2003epigenetic,rohlf2012modeling}. These modifications can be fixed to various levels, via which, gene expression levels can be fixed continuously, rather than at a discrete set of values as in attractors. \tcb{As will be shown, the expression level, in terms of dynamical systems, are not given as a fixed-point attractor but are given as a continuous state. Now, the growth rate depends on the expression level, and cellular states with different growth rates are generated depending on modification levels. \\}
\tcb{In the present Letter, we demonstrate that adaptation to achieve faster growth for given environmental conditions is generally attained only by noise and the growth-induced dilution, without preparing specific gene expression networks with multiple attractors. We will provide a general concept of selection from and discuss their relevance to cell biology.\\
Here, we study a simple cell model consisting of proteins and the degree of epigenetic modification levels of genes. In the model, we take into account the following processes: (i) Each protein is synthesized at a certain rate according to the expression level of genes; (ii) proteins are spontaneously degraded; (iii) Some specific proteins determine the growth rate of the cell, whereas proteins are diluted due to the volume growth of the cell \tcb{(the cell division is incorporated in the dilution process)}. (iv) For specific genes, epigenetic modifications are taken into account and can alter the expression levels of genes, whereas the modification changes need specific molecular machinery.\\}
\tcb{
Now, we introduce a simple model consisting of the expression level of two types of genes and the corresponding proteins: target ($t$) and sensor ($s$) genes and proteins\footnote{Here, each expression variable does not necessarily represent a single gene or protein but can be regarded as a collective variable representing the average behavior of genes or proteins that share similar functions.}. \\ 
The target protein contributes to the growth of the cell, whose rate $\mu$, thus, $\mu$ is the function of its concentration. As the simplest example, we suppose that the synthesis rate of the target protein is determined by an epigenetic modification level, while the sensor protein is produced in a constant rate, $v_s$.\\
Although the molecular mechanisms that regulate the epigenetic modification have not yet been fully elucidated, it is natural to assume that the epigenetic modification is regulated some other genes, which we call here as the sensor gene. If the concentration of the sensor protein exceeds a threshold value, it facilitates the synthesis of some proteins that modify the epigenetic state of the target gene. By means of the rate equations, the dynamics of the above model are given as}
\tcb{
\begin{eqnarray}
 \frac{d E_t}{dt}&=&H(P_s)(l_+ - l_-E_t),  \label{eq:Et}\\
 \frac{dP_t}{dt}&=&v_tE_t - d_tP_t - \mu P_t,\label{eq:Pt}\\
  \frac{dP_s}{dt}&=&v_s - d_sP_s - \mu P_s,\label{eq:Ps}
 \end{eqnarray} with 
where $E_t$, $P_t$, and $P_s$ represents the level of the epigenetic modification, the concentration of the target protein, and the concentration of the sensor protein, respectively. $P_t$ is produced in proportional to the epigenetic modification level, with a rate $v_tE_t$. $d_s$ and $d_t$ represent the degradation rate of the corresponding protein. $l_+$ and $l_-$ is the activation and inhibition rate of the epigenetic state $E_t$, respectively.\\
The modification by the sensor protein is represented by $H(P_s)$ given by a sigmoidal function, $H(P_s)=1/(1+\exp(-\beta(P_s-\theta)))$, where $\theta$ is the threshold for the induction, and $\beta$ is steepness of this increase, corresponding to the Hill coefficient\footnote{\tcb{If $H$ is replaced by a Hill function, the results are not altered qualitatively. We showed the result with a Hill function in {\it Supplement}.}}}. This form is adopted as a natural consequence of binding--unbinding kinetics between a DNA molecule and transcription factors, with the inclusion of the effect of cooperativity. \tcb{The target protein is synthesized at the rate proportional to the epigenetic state of the gene\footnote{\tcb{In the equations (\ref{eq:Pt})-(\ref{eq:Ps}), a gene and corresponding protein to modify the epigenetic state are not explicitly included to the model for simplicity's sake. We have studied also a model in which the modifier gene and protein are implemented, and have confirmed that the simplification is not necessary to obtain the main results. and have confirmed that the results are qualitatively same with what shown in the main text.}}}. Here we set specific growth rate $\mu$ to $\mu=\mu_{\rm max}/(1+(P_t-P_t^*)^2)$ (i.e., there is an optimal concentration of the target protein to maintain the cell growth). The results shown below, however, do not change qualitatively if we choose other forms of $\mu$ (see {\it Supplement}). \tcb{Note that any specific input to increase the growth rate is not introduced here.}\\
To study the influence of stochasticity on chemical reactions, we employed stochastic simulations \tcb{of the reactions corresponding to Eq.(\ref{eq:Et})-(\ref{eq:Ps}), via the Gillespie algorithm\cite{gillespie1977exact} (i.e., in the continuous limit with a large number of molecules Eq.(\ref{eq:Et})-(\ref{eq:Ps}) is regained). The master equation which we used for the simulation is shown in the {\it Supplement}. \\}
In Fig.\ref{fig:simple}(a), examples of the time course of the growth rates are plotted for two different values of threshold $\theta$, where the initial value of $E_t$ and $P_t$ at the steady-state values of Eq.(\ref{eq:Et})-(\ref{eq:Ps})  with $\theta=0$, so that the growth rate is low initially. The time courses show that the growth rate of a cell switches to a high value when $\theta$ is large, whereas for small $\theta$ (black curve), it remains at the initial low value. In the former case, the epigenetic state of the target gene is modified so that the cell achieves a high growth rate \footnote{Note that even if we start from a fast-growth state, the cell cannot maintain the fast growth when $\theta=0$. Thus, the failure of this adaptation at $\theta=0$ is not due to the choice of the initial state.}. Dependence of the averaged growth rate on $\theta$ is plotted in Fig.~\ref{fig:simple}(b), which shows that the growth rate starts to rise at approximately $\theta \approx 150$ and reaches the maximal value at $\theta \approx 200$.\\
When we set $\theta$ and the noise level appropriately, the adaptation mechanism works. It is intuitively explained as follows: If the growth rate of the cell is low (i.e., $P_t$ is far from $P_t^*$), then the sensor protein accumulates and $P_s$ exceeds the threshold value for $H(P_s)$, $\theta$, and then the epigenetic state of the target gene is modified. If the epigenetic state of a target gene reaches an appropriate value to allow for the increase in the growth rate, then all proteins are strongly diluted accordingly. Then, the concentration of the sensor protein hardly exceeds the threshold value because of the strong dilution by rapid growth, so that the modification of the epigenetic variable takes place at a low probability. As shown in Fig.~\ref{fig:simple}(c), the probability of the change in the epigenetic variable suddenly drops to almost zero when $\mu$ is increased. Thus, the epigenetic state of the target gene remains at a certain value which supports high-growth rate. If the cell remains at a slow-growth state, however, then the epigenetic modification state keeps on being changed by noise. Hence, the system sooner or later reaches a fast-growth state and stays there.\\
\tcb{Three conditions are essential for this adaptation mechanism to the fast-growth state work. \\
\tcb{(i).{\bf the noise level}}, due to the finiteness of the molecule number in a cell. With the decrease in the number, the noise level due to a stochastic reaction process is increased. In Fig.~\ref{fig:simple}(d), the average growth rate is plotted against the average number of molecules. If the number is too small, noise dominates over the average gene expression dynamics, so that the cellular state fluctuates almost randomly. On the other hand, when the number is too large, the noise level is so small that the state is hardly kicked out from the initial low-growth attractor of the deterministic equation (Eq.(\ref{eq:Et})-(\ref{eq:Ps})). At an intermediate noise level, the noise is dominant for a slow-growth state, and once a fast-growth state is reached by the fluctuation, the deterministic change via Eq.(\ref{eq:Et})-(\ref{eq:Ps}) dominates over the stochasticity, and the state remains therein. Thus, selection of a fast-growth state then occurs. \\
\tcb{(ii). {\bf "line attractor".}} Here, the reached states with high growth rates are not given as a unique fixed point but rather on a line of marginal stable states (termed as line attaractor). The state can take any value on the line continuously, whereas it is attracted across the line: Recalling that the dynamics of the averaged concentration of the target molecule is written as $\dot{E}_t=H(P_s)(l_+-l_-E_t)$, $E_t$ relaxes to $l_+/l_-$ if $P_s$ constantly exceeds the threshold ($H\neq 0$). Here, however, $H$ becomes vanishingly small if $P_s$ is far below the threshold. Then, $E_t$ can stay at an arbitrary value. Hence the points along the line with arbitrary value of $E_t$ and $H(P_s)=0$ are marginally stable attractors. \\
\tcb{(iii). {\bf timescale separation between the epigenetic variable and other variables}}. First of all, the change in the $E_t$ value must be quickly reflected in the concentration of the target protein concentration $P_t$. If the dynamics of $P_t$ are considerably slower than those of $E_t$, then $P_t$ would not reach the value giving rise to the high growth rate, even if $E_t$ reached the value corresponding to the fast growth. Additionally, the dynamics of $P_s$ have to be faster than those of $E_t$. Otherwise, the reached steady-state value of $E_t$ cannot afford a high growth rate.\\
 From Eq.(\ref{eq:Et})-(\ref{eq:Ps}), $E_t$ and $P_n\ (n=s,t)$  exponentially relaxes to its steady state solution with the rate approximately $l_-$, and $d_n+\mu_{\rm st}$, respectively, where $\mu_{\rm st}$ is the mean growth rate near the steady state. Hence, the relation $l_-\ll d_n +\mu_{\rm st}$ is the condition for the adaptation mechanism to work.\\}
\tcb{
Noting this time-scale difference, we can analyze the conditions for this adaptive growth, by eliminating $P_t$ and $P_s$ adiabatically and deriving the Fokker--Plank equation of $E_t$, under the assumption that the number of the proteins are sufficiently large. In addition, for the ease of analysis, we study the cases in which $d_t\gg\mu_{\rm max}$ holds. \tcb{Under these approximations, we obtain the Fokker--Planck equation for the epigenetic variable given by
\begin{eqnarray}  
\frac{\partial }{\partial t}F(E_t,t)&=&-\frac{\partial}{\partial E_t}\tilde{H}(E_t)(l_+-l_-E_t)F(E_t,t)\nonumber \\&+&\frac{\sigma^2}{2}\frac{\partial^2}{\partial E_t^2}\tilde{H}(E_t)(l_++l_-E_t)F(E_t,t)\label{eq:reduce_det}\nonumber \\
\tilde{H}(E_t)&=&\frac{1}{1+\exp[-\ti{\beta}((E_t-E_t^*)^2-\ti{\theta})]}
\end{eqnarray}  
where $F(E_t,t)$ is the probability density function that the epigenetic variable has the value $E_t$ at time $t$. The parameters are given by $\ti{\beta}=\beta(v_t/d_t)^2  v_s/\mu_{\rm max}$, $\ti{\theta}=(\theta  \mu_{\rm max}/v_s-1)(d_t/v_t)^2$, and $E_t^*=P_t^* d_t/v_t$, while $\sigma$ represents the noise stregth (Details of the derivation of Eq.(\ref{eq:reduce_det}) and following calculations are given in {\it Suppelement}}). The steady solution $F_{\rm st}(E_t)$ is obtained, as is plotted for several values of $\theta$ in Fig.~\ref{fig:FP}(a). The peak of the steady distribution shows a transition from the trivial value corresponding to the stable fixed point value of $E_t$ in the noiseless limit to the adaptive value $E_t\sim E_t^*$.  The average growth rate thus obtained is plotted as a function of $\sigma$ and $\theta$ in Fig.\ref{fig:FP}(b). Accordingly, \tcb{a boundary which separates the adaptive and non-adaptive regimes is analytically estimated (see {\it Supplement}). The estimate well captures the transition between two regimes \footnote{\tcb{The growth rate remains at the high values even if $\sigma$ increases further as long as $\theta$ is sufficiently large, whereas in the infinitely large noise region, our assumption that the effect of the stochasticity can be neglected for $P_s$ and $P_t$ is no longer valid.}}.} \\ 
To verify the robustness of the presented results, 
we confirmed that the adaptation mechanism works independently of the specific form of the growth rate and parameter values. Furthermore, even if the epigenetic modification state changes spontaneously, a fast-growth state is selected and sustained, as long as \tcb{the modifications triggered by the sensor protein} dominates the epigenetic modification. The detailed set-ups and the results are presented in {\it Supplement}.\\
In addition, we tested the reliability of our adaptation mechanism in chemical reaction network systems in which metabolites are converted into growth factors via several enzymatic reactions. \tcb{In the model, the epigenetic variable is implemented to each enzyme, whereas one sensor protein triggers the changes of all the epigenetic variables. As shown in Fig.\ref{fig:network}, the proposed mechanism works also in the chemical reaction network model\footnote{\tcb{The achieved growth rate here is still lower than $\mu_{\rm max}$ (e.g., 20\% of it), the possible maximal value. Recall, however, that the initially the growth rate is $\sim 0$, and it is enhanced in the orders of magnitude, just by the fluctuations in the epigenetic modiﬁcation level. The growth rate is then sustained at a high level. Note also that just a random reaction network is chosen here, without tuning in the parameter values, either. Optimizing parameter values or network structures would further increase the growth rate.}}. Details of the model is given in the {\it Supplement}.}  \\}

\tcb{To sum up, numerical simulations and analytic calculation of our model revealed that the epigenetic regulation and stochasticity of chemical reactions lead to an increase in the growth rate of the cell, starting from the stable attractor with a low growth rate. Even though there is no predesigned gene-regulation or signal transduction network to achieve the high growth rate, the cell reaches a high-growth state, and stays therein driven by stochastic noise and dilution by cell-volume increase.}\\
If the growth rate is high, the dilution is large, so that the protein concentrations relax fast, and the perturbation by noise has little influence. For the slow-growth state, the protein concentrations fluctuate more under the action of noise, whereas the epigenetic state is frequently modified due to the accumulation of the sensor protein, which enables reaching a fast-growth state. \tcb{This directional epigenetic change driven by noise may be reminiscent of thermal ratchet \cite{vale1990protein,feynman2011feynman,julicher1995cooperative,magnasco1993forced,julicher1997modeling}, but here it is due to noise in the reaction and epigenetic process. Given that this ``epigenetic ratchet" does not require any tuned design by evolution, it can explain the robust adaptation observed experimentally \cite{kashiwagi2006adaptive,kotte2010bacterial,braun2015unforeseen}.\\}
The importance of the epigenetic regulatory mechanisms has been confirmed both for prokaryotes and eukaryotes by a number of studies \cite{casadesus2006epigenetic,croken2012chromatin,rovira2012transcriptional}. It is also reported theoretically that the epigenetic modifications allow for compatibility between evolutional robustness and plasticity \cite{gombar2014epigenetics}. Although factors that provoke epigenetic modification events remain to be elucidated, some evidence suggests that epigenetic modifications are incuded by environmental changes.\cite{rovira2012transcriptional}. Besides, the epigenetic modification contains noise because of the inherent stochasticity of gene expression levels and chemical reactions. \tcb{Indeed, switches between the two epigenetic states \tcb{in the $lac$ systems} are reported to be stochastic \cite{choi2008stochastic,satory2011epigenetic}. }Furthermore, epigenetic modifications are inherited over generations as ``epigenetic memory" \cite{ronin2017long}, which may indicate slow dynamics of epigenetic modifications. These experimental reports support the assumptions on the epigenetic modifications required in the present study. \\
\tcb{Recently, Rocco et. al.\cite{rocco2013slow} proposed a possible scenario for the bacterial persistence \cite{dhar2007microbial,balaban2004bacterial}, based on the transcriptional noise and slow relaxation of the protein concentration (instead of the epigenetic change in our model). We hope that the adaptive mechanism proposed here could also shed light on persistence.}\\
Note that the three adaptation mechanisms, i.e., designed signal transduction \cite{jacob1961genetic}, attractor selection \cite{kashiwagi2006adaptive,furusawa2008generic,furusawa2013epigenetic}, and the present epigenetic mechanism are not mutually exclusive. Using these three adaptation mechanisms properly according to the situations, cells may realize the flexible adaptation.\\
The authors would like to thank K. Sneppen and N. Mitarai for useful discussions. This research was partially supported by a Grant-in-Aid for Scientific Research (S) (15H05746),
Grant-in-Aid for Scientific Research on Innovative Areas (17H06386) from the Ministry of Education, Culture, Sports, Science and Technology (MEXT) of Japan, and JSPS Grant No. 16J10031. \\


\begin{thebibliography}{10}

\bibitem{jacob1961genetic}
Fran{\c{c}}ois Jacob and Jacques Monod.
\newblock Genetic regulatory mechanisms in the synthesis of proteins.
\newblock {\em Journal of molecular biology}, 3(3):318--356, 1961.

\bibitem{kotte2010bacterial}
Oliver Kotte, Judith~B Zaugg, and Matthias Heinemann.
\newblock Bacterial adaptation through distributed sensing of metabolic fluxes.
\newblock {\em Molecular systems biology}, 6(1):355, 2010.

\bibitem{braun2015unforeseen}
Erez Braun.
\newblock The unforeseen challenge: from genotype-to-phenotype in cell
  populations.
\newblock {\em Reports on Progress in Physics}, 78(3):036602, 2015.

\bibitem{kashiwagi2006adaptive}
Akiko Kashiwagi, Itaru Urabe, Kunihiko Kaneko, and Tetsuya Yomo.
\newblock Adaptive response of a gene network to environmental changes by
  fitness-induced attractor selection.
\newblock {\em PloS one}, 1(1):e49, 2006.

\bibitem{stern2007genome}
Shay Stern, Tali Dror, Elad Stolovicki, Naama Brenner, and Erez Braun.
\newblock Genome-wide transcriptional plasticity underlies cellular adaptation
  to novel challenge.
\newblock {\em Molecular Systems Biology}, 3(1):106, 2007.

\bibitem{stolovicki2006synthetic}
Elad Stolovicki, Tali Dror, Naama Brenner, and Erez Braun.
\newblock Synthetic gene-recruitment reveals adaptive reprogramming of gene
  regulation in yeast.
\newblock {\em Genetics}, 2006.

\bibitem{tsuru2011adaptation}
Saburo Tsuru, Nao Yasuda, Yoshie Murakami, Junya Ushioda, Akiko Kashiwagi,
  Shingo Suzuki, Kotaro Mori, Bei-Wen Ying, and Tetsuya Yomo.
\newblock Adaptation by stochastic switching of a monostable genetic circuit in
  escherichia coli.
\newblock {\em Molecular systems biology}, 7(1):493, 2011.

\bibitem{shimizu2011stochastic}
Yoshihiro Shimizu, Saburo Tsuru, Yoichiro Ito, Bei-Wen Ying, and Tetsuya Yomo.
\newblock Stochastic switching induced adaptation in a starved escherichia coli
  population.
\newblock {\em PLoS One}, 6(9):e23953, 2011.

\bibitem{tan2009emergent}
Cheemeng Tan, Philippe Marguet, and Lingchong You.
\newblock Emergent bistability by a growth-modulating positive feedback
  circuit.
\newblock {\em Nature chemical biology}, 5(11):842, 2009.

\bibitem{furusawa2008generic}
Chikara Furusawa and Kunihiko Kaneko.
\newblock A generic mechanism for adaptive growth rate regulation.
\newblock {\em PLoS Comput Biol}, 4(1):e3, 2008.

\bibitem{furusawa2013epigenetic}
Chikara Furusawa and Kunihiko Kaneko.
\newblock Epigenetic feedback regulation accelerates adaptation and evolution.
\newblock {\em PloS one}, 8(5):e61251, 2013.

\bibitem{jaenisch2003epigenetic}
Rudolf Jaenisch and Adrian Bird.
\newblock Epigenetic regulation of gene expression: how the genome integrates
  intrinsic and environmental signals.
\newblock {\em Nature genetics}, 33:245, 2003.

\bibitem{rohlf2012modeling}
Thimo Rohlf, Lydia Steiner, Jens Przybilla, Sonja Prohaska, Hans Binder, and
  J{\"o}rg Galle.
\newblock Modeling the dynamic epigenome: from histone modifications towards
  self-organizing chromatin.
\newblock {\em Epigenomics}, 4(2):205--219, 2012.

\bibitem{Note1}
Here, each expression variable does not necessarily represent a single gene or
  protein but can be regarded as a collective variable representing the average
  behavior of genes or proteins that share similar functions.

\bibitem{Note2}
\protect \leavevmode {\protect \color {black}If $H$ is replaced by a Hill
  function, the results are not altered qualitatively. We showed the result
  with a Hill function in {\protect \it Supplement}.}

\bibitem{Note3}
\protect \leavevmode {\protect \color {black}In the equations (\ref
  {eq:Pt})-(\ref {eq:Ps}), a gene and corresponding protein to modify the
  epigenetic state are not explicitly included to the model for simplicity's
  sake. We have studied also a model in which the modifier gene and protein are
  implemented, and have confirmed that the simplification is not necessary to
  obtain the main results. and have confirmed that the results are
  qualitatively same with what shown in the main text.}

\bibitem{gillespie1977exact}
Daniel~T Gillespie.
\newblock Exact stochastic simulation of coupled chemical reactions.
\newblock {\em The journal of physical chemistry}, 81(25):2340--2361, 1977.

\bibitem{Note4}
Note that even if we start from a fast-growth state, the cell cannot maintain
  the fast growth when $\theta =0$. Thus, the failure of this adaptation at
  $\theta =0$ is not due to the choice of the initial state.

\bibitem{Note5}
\protect \leavevmode {\protect \color {black}The growth rate remains at the high
  values even if $\sigma $ increases further as long as $\theta $ is
  sufficiently large, whereas in the infinitely large noise region, our
  assumption that the effect of the stochasticity can be neglected for $P_s$
  and $P_t$ is no longer valid.}

\bibitem{Note6}
\protect \leavevmode {\protect \color {black}The achieved growth rate here is
  still lower than $\mu _{\protect \rm max}$ (e.g., 20\% of it), the possible
  maximal value. Recall, however, that the initially the growth rate is $\sim
  0$, and it is enhanced in the orders of magnitude, just by the fluctuations
  in the epigenetic modiﬁcation level. The growth rate is then sustained at a
  high level. Note also that just a random reaction network is chosen here,
  without tuning in the parameter values, either. Optimizing parameter values
  or network structures would further increase the growth rate.}

\bibitem{vale1990protein}
Ronald~D Vale and Fumio Oosawa.
\newblock Protein motors and maxwell's demons: does mechanochemical
  transduction involve a thermal ratchet?
\newblock {\em Advances in biophysics}, 26:97--134, 1990.

\bibitem{feynman2011feynman}
Richard~P Feynman, Robert~B Leighton, and Matthew Sands.
\newblock {\em The Feynman lectures on physics, Vol. I: The new millennium
  edition: mainly mechanics, radiation, and heat}, volume~1.
\newblock Basic books, 2011.

\bibitem{julicher1995cooperative}
Frank J{\"u}licher and Jacques Prost.
\newblock Cooperative molecular motors.
\newblock {\em Physical review letters}, 75(13):2618, 1995.

\bibitem{magnasco1993forced}
Marcelo~O Magnasco.
\newblock Forced thermal ratchets.
\newblock {\em Physical Review Letters}, 71(10):1477, 1993.

\bibitem{julicher1997modeling}
Frank J{\"u}licher, Armand Ajdari, and Jacques Prost.
\newblock Modeling molecular motors.
\newblock {\em Reviews of Modern Physics}, 69(4):1269, 1997.

\bibitem{casadesus2006epigenetic}
Josep Casades{\'u}s and David Low.
\newblock Epigenetic gene regulation in the bacterial world.
\newblock {\em Microbiology and molecular biology reviews}, 70(3):830--856,
  2006.

\bibitem{croken2012chromatin}
Matthew~M Croken, Sheila~C Nardelli, and Kami Kim.
\newblock Chromatin modifications, epigenetics, and how protozoan parasites
  regulate their lives.
\newblock {\em Trends in parasitology}, 28(5):202--213, 2012.

\bibitem{rovira2012transcriptional}
N{\'u}ria Rovira-Graells, Archna~P Gupta, Evarist Planet, Valerie~M Crowley,
  Sachel Mok, Llu{\'\i}s~Ribas de~Pouplana, Peter~R Preiser, Zbynek Bozdech,
  and Alfred Cort{\'e}s.
\newblock Transcriptional variation in the malaria parasite plasmodium
  falciparum.
\newblock {\em Genome research}, 22(5):925--938, 2012.

\bibitem{gombar2014epigenetics}
Saurabh Gombar, Thomas MacCarthy, and Aviv Bergman.
\newblock Epigenetics decouples mutational from environmental robustness. did
  it also facilitate multicellularity?
\newblock {\em PLoS computational biology}, 10(3):e1003450, 2014.

\bibitem{choi2008stochastic}
Paul~J Choi, Long Cai, Kirsten Frieda, and X~Sunney Xie.
\newblock A stochastic single-molecule event triggers phenotype switching of a
  bacterial cell.
\newblock {\em Science}, 322(5900):442--446, 2008.

\bibitem{satory2011epigenetic}
Dominik Satory, Alasdair~JE Gordon, Jennifer~A Halliday, and Christophe Herman.
\newblock Epigenetic switches: can infidelity govern fate in microbes?
\newblock {\em Current opinion in microbiology}, 14(2):212--217, 2011.

\bibitem{ronin2017long}
Irine Ronin, Naama Katsowich, Ilan Rosenshine, and Nathalie~Q Balaban.
\newblock A long-term epigenetic memory switch controls bacterial virulence
  bimodality.
\newblock {\em Elife}, 6, 2017.

\bibitem{rocco2013slow}
Andrea Rocco, Andrzej~M Kierzek, and Johnjoe McFadden.
\newblock Slow protein fluctuations explain the emergence of growth phenotypes
  and persistence in clonal bacterial populations.
\newblock {\em PloS one}, 8(1):e54272, 2013.

\bibitem{dhar2007microbial}
Neeraj Dhar and John~D McKinney.
\newblock Microbial phenotypic heterogeneity and antibiotic tolerance.
\newblock {\em Current opinion in microbiology}, 10(1):30--38, 2007.

\bibitem{balaban2004bacterial}
Nathalie~Q Balaban, Jack Merrin, Remy Chait, Lukasz Kowalik, and Stanislas
  Leibler.
\newblock Bacterial persistence as a phenotypic switch.
\newblock {\em Science}, 305(5690):1622--1625, 2004.

\end{thebibliography}

\end{document}